\documentstyle[preprint,aps]{revtex}
\begin{document}
\newcommand\ie {{\it i.e.}}
\newcommand\eg {{\it e.g.}}
\newcommand\etc{{\it etc.}}
\newcommand\cf {{\it cf.}}
\newcommand\etal {{\it et al.}}
\newcommand{\be}{\begin{eqnarray}}
\newcommand{\ee}{\end{eqnarray}}
\newcommand{\jp}{$ J/ \psi $}
\newcommand{\up}{$ \Upsilon$}
\newcommand{\pp}{$ \psi^{ \prime} $}
\newcommand{\ppp}{$ \psi^{ \prime \prime } $}
\newcommand{\dd}[2]{$ #1 \overline #2 $}
\newcommand\noi {\noindent}
\newcommand\xb  {$ x_{Bj}$}
\draft
\preprint{LBL-35821}
\title
{Consequences of Nuclear Shadowing for Heavy Quarkonium
Production in Hadron-Nucleus Interactions$^{\star}$}
\footnotetext{$\star$ This work was supported in part by the Director,
Office of Energy
Research, Division of Nuclear Physics of the Office of High Energy and Nuclear
Physics of the U. S. Department of Energy under
Contract Number DE-AC03-76SF0098.}

\author{S. Liuti}
\address
{Nuclear Science Division, Lawrence Berkeley Laboratory, Berkeley CA 94720,
USA \\
and
\\
I.N.F.N. Sezione Sanit\'{a} \\
Viale Regina Elena 299, I-00161 Rome, Italy}
\author{R. Vogt}
\address
{Nuclear Science Division, Lawrence Berkeley Laboratory, Berkeley CA 94720,
USA}
\maketitle

\begin{abstract}
We study nuclear shadowing in $J/\psi$ and $\Upsilon$
production in hadron-nucleus interactions
and in nucleus-nucleus collisions at the Relativistic Heavy Ion Collider
and the Large Hadron Collider.
As a consequence of the perturbative $Q^2$-dependence of gluon shadowing,
we predict that $\Upsilon$ production is less suppressed
than the $J/\psi$.
We show that antishadowing leads
to enhanced \jp\ production at $x_f \lesssim 0$, an effect reduced for
$\Upsilon$ production.
\end{abstract}
\newpage

\vspace{1.0 cm}
The production of charmonium and bottomonium states in hadron-nucleus
interactions does not increase linearly with the nuclear target size $A$.
Additionally, the ratio of hadron-nucleus to hadron-nucleon
production, $S_A = \sigma_{hA}/A\sigma_{hp}$, is not constant but
decreases as a
function of the fraction of the center of mass momentum, $x_f$, carried by
the produced
resonance.  There are several effects that contribute both to the $A$
dependence of the total cross section and to the ratio $S_A$
including nuclear absorption \cite{Gerschel},
comover interactions \cite{GavinVogt},
intrinsic heavy quark states \cite{VBH1}, projectile energy loss
\cite{GavinMilana}, and nuclear shadowing \cite{MuellerQiu,GuptaSatz}.
%
%

In this paper, we examine the role of shadowing of the nuclear parton
distributions in quarkonium production.
Nuclear shadowing modifies the target parton distributions
so that $xq^A(x) \neq A xq^N(x)$.
Deep inelastic scattering (DIS) data
on the ratio $R_A=F_2^A/F_2^D$
\cite{Arneodo,E665} show that:
{\em i)} nuclear shadowing begins to set in at
$x_{Bj} \lesssim 0.06$ for all nuclei ($R_A <1$);
{\em ii)} $R_A$ has a very weak $Q^2$ dependence; and {\em iii)}
an enhancement, or {\em antishadowing}, exists for $0.06 \lesssim x_{Bj}
\lesssim 0.25$  ($R_A > 1$).
In addition, the NMC data on $J/\psi$ production
with Sn and C targets show an enhancement of $(13 \pm 8)\%$ for
$0.05 \lesssim x_{Bj} \lesssim 0.3$ \cite{NMC},
interpreted as gluon antishadowing.

The weak $Q^2$ dependence of $R_A$ is consistent
with the models of \cite{NikZak,FSPRep}, where
nuclear shadowing is considered as a leading-order
QCD effect with perturbatively generated $Q^2$ dependence.
These models differ in the treatment of the nonperturbative contribution
at the initial scale $Q^2_0$.  We adopt
the QCD Aligned-Jet Model (QAJM) \cite{FSPRep} which
accounts for the DIS data \cite{FSPRep,FSLiuti} to study the effect of nuclear
shadowing on $S_A$.

We would like to point out two major consequences
of shadowing and antishadowing for \jp\ and \up\ hadroproduction based on very
general arguments.
Since $m_\Upsilon$ is three times larger than $m_{\psi}$, the $\Upsilon$
is primarily produced within the antishadowing region of the target
at present energies.
However, the parton distributions are evaluated at $Q^2 = m_\Upsilon^2$, thus
$Q^2$ evolution reduces the effect of both shadowing
and antishadowing in \up\ production with respect to \jp.
The antishadowing apparently present in both the
DIS measurements of $R_A$ \cite{Arneodo,E665}
and the gluon distribution \cite{NMC} reflects
on $S_A$.  We include the observed antishadowing effect, a general consequence
of baryon number and momentum conservation in a nucleus, and
explore the consequences of $Q^2$ evolution on $S_A$.

We do not expect our shadowing model to fully explain the
shape and magnitude of $S_A$.  To
quantitatively show the effects of nuclear
shadowing, we have not included other $A$ dependent
effects.  Nuclear absorption and comover
interactions do not have a strong $x_f$
dependence and mainly affect the magnitude of $S_A$
\cite{Gerschel,GavinVogt}.
On the other hand, intrinsic heavy quarks \cite{VBH1} and
projectile energy loss
\cite{GavinMilana} have a significant $x_f$ dependence.
%
%

The leading order \dd{Q}{Q} bound state cross section is the integral of
the free \dd{Q}{Q} production cross section from the \dd{Q}{Q}
threshold to the meson pair threshold,
\be \frac{d \sigma}{d x_f} = 2F
\int_{2m_Q/\sqrt{s}}^{2m_H/\sqrt{s}}
\tau d\tau \frac{H_{pt}(x_1,x_2;x_1 x_2 s)}{\sqrt{x_f^2
+ 4 \tau^2}} \, \, , \ee
where $m_c = 1.5$ GeV, $m_b = 4.75$ GeV, $m_D = 1.867$ GeV, and
$m_B = 5.28$ GeV.  The bound state fraction, $F$, is
a parameter that cancels in $S_A$.  We include
quark--antiquark annihilation and gluon--gluon fusion
subprocesses in the convolution formula $H_{pt}$,
\be
        H_{pt}(x_1,x_2; m^2) & = & G_p(x_1)G_t(x_2) \sigma(gg \rightarrow Q
\overline Q;m^2)        \nonumber \\                         &   & \mbox{} +
\sum_{q=u,d,s} (q_p(x_1) \overline q_t(x_2) + \overline q_p(x_1)
q_t(x_2)) \sigma(q \overline q \rightarrow Q \overline Q;m^2) \, \, ,
\label{Hpt}
\ee
where $x_{1,2} = \frac{1}{2}
(\pm x_f + \sqrt{x_f^2 + 4 \tau^2})$ are the projectile and target momentum
fractions.  We use the MRS D$-^\prime$ \cite{D0} distributions that describe
the recent HERA data on $F_2^{ep}$ \cite{HERA}.

Next-to-leading order contributions to the free $Q \overline Q$
cross section
have been
calculated.  The $x_f$ distribution is approximately that from lowest order
multiplied by a theoretical $K$ factor,
$\sigma(\alpha_s^3)/\sigma(\alpha_s^2)$ \cite{MNR}, assumed to be absorbed
in $F$.  At lowest order, the \jp\ is produced
with $p_T=0$ since the $c \overline c$ pair is back-to-back.
A $p_T$ dependence may be introduced through an intrinsic
transverse momenta, $q_T$, of the initial partons as we show later.
For $p_T \gg m_{\rm res}$, higher-order
corrections are needed \cite{EllisSexton}.  Since we concentrate on the $x_f$
distributions and the $A$ dependence, where $p_T \lesssim
m_{\rm res}$ is dominant, our calculation is lowest order.

When the target is a nucleus, the parton distributions,
$G_t(x_2)$, $q_t(x_2)$ and $\bar{q}_t(x_2)$ in Eq.\ (\ref{Hpt}) are
the nuclear medium modified 
distributions evaluated at $Q^2 \approx m_{\psi}^2$ and $Q^2 \approx
m_{\Upsilon}^2$ for \jp\ and \up\ production.  We calculate the shadowed
distributions at the initial scale, $Q^2=Q^2_0 \approx {\rm 5 \, GeV^2}$,
and $x<x_{sh} \approx 0.04$
according to the QAJM \cite{FSPRep} and evolve in $Q^2$ using the
Dokshitser-Gribov-Lipatov-Altarelli-Parisi equations (DGLAP) \cite{GDLAP}.
Recombination effects within a single nucleon are negligible at
$x \gtrsim 10^{-3}$ \cite{MuellerQiu,EQW}.
In the QAJM, shadowing is generated
by the interaction of hadronic configurations
originating from fluctuations of the virtual photon
with small transverse momentum, or
jets {\em aligned} along the direction of the virtual photon,
$\vec{{\bf q}}/| {\bf q}|$,
with $k_\perp < k_{0 \perp } \approx 0.4 \, {\rm GeV}$.
The transverse separation  of the aligned jets is correspondingly
large ($\approx$ 1 fm) and their phase space
is restricted by a factor $ \propto k_{0 \perp}^2/M^2$.
In QCD the transverse momentum of the
fluctuation can be large so that
{\em nonaligned} hadronic configurations are simultaneously present.
Such configurations interact differently in the nuclear medium.
In Refs.\ \cite{FSPRep,FSLiuti} it was assumed that aligned
jets with large transverse separation interact in the nucleus as
vector mesons. Nonaligned jets, with $k_\perp > k_{0 \perp}$ and a
small transverse separation have a correspondingly small interaction
cross section, giving rise to color transparency.

The nuclear baryon number
and momentum sum rules:
\begin{mathletters}
\begin{eqnarray}
\lefteqn{\frac{1}{A} \int_0^A dx V_A(x,Q^2) = \int_0^1 dx V_N(x,Q^2)
\, \, ,}  \\
& & \frac{1}{A} \int_0^A dx x [V_A(x,Q^2)+G_A(x,Q^2)+S_A(x,Q^2)]= \nonumber\\
& & \int_0^1 dx x [V_N(x,Q^2)+G_N(x,Q^2)+S_N(x,Q^2)] \, \, ,
\label{sumrules}
\end{eqnarray}
\end{mathletters}
\noindent
must be fulfilled at every $Q^2$.  In Eq.\ (\ref{sumrules})
$V_{N(A)} \equiv u_t^V + d_t^V$,
$S_{N(A)} \equiv 2(\overline u_t + \overline d_t + \overline s_t)$, and
$G_{N(A)} \equiv G_t$, are the nucleon, $N$,
and nuclear, $A$, valence, sea and gluon distributions.
In order to satisfy the sum rules and simultaneously allow for
shadowing at $x \lesssim 0.04$, the parton distributions
must be enhanced at higher values of $x$.
We assume that the enhancement is shared by the valence
quarks and gluons since the sea quark enhancement
is consistent with zero \cite{E772} (see also Ref.\ \cite{FSLiuti}).
Recent data from E665 \cite{E665} show that the
enhancement in $R_A$ is concentrated in the interval
$0.06 \lesssim x \lesssim 0.25$ and is $A$-independent.  These
two crossover points, assuming no sea
enhancement, are sufficient to constrain the valence and gluon antishadowing.

Gluon shadowing should be larger than sea quark shadowing since the
interaction of a $q\overline q g$ configuration in the
$\gamma ^{\ast}$ wavefunction with the 
target should be stronger than a $q \overline q$
configuration \cite{FSLMoriond,NikZak93}.
Indeed, the perturbative interaction cross section of
a small color octet with a nucleon is $9/4$ larger than a color
triplet of the same transverse size.
Assuming a smooth connection between the perturbative and
nonperturbative domains,
$\sigma_{\gamma^* \rightarrow q \overline
qg,N} = \frac{9}{4} \sigma_{\gamma^* \rightarrow q\overline q,N}$,
leading to larger effects on the gluon
relative to the sea and valence quarks \cite{FSLMoriond}.
The overall enhancement found in \cite{FSLiuti,FSLMoriond}
is consistent with the data of \cite{NMC},
used to constrain the shape of the gluon distribution in the
antishadowing region.
A weak $Q^2$ dependence for $R_A$ is obtained, in accord with the data of
\cite{Arneodo,E665}, while a stronger $Q^2$ dependence is found for the
gluon ratio, $R^G_A=G_A/G_N$. Similar results were also obtained
in \cite{Eskola}.

In Fig.\ 1 we show $x_2$ as a function of $x_f$ for \jp\ (solid curve),
\up\ (dashed) production for several energies
to point out the shadowing and antishadowing regions.
At 200 GeV, 1(a), the \jp\ is antishadowed for $0 \lesssim x_f \lesssim 0.6$.
At 800 GeV, 1(b), where data exists for \jp\ and \up\ from $-0.2 <x_f<0.65$
\cite{E772q}, the \jp\ lies in the shadowing region for $x_f>0$ but enters
the antishadowing region at backward values.  The \up\ is mainly produced in
the antishadowing
region, but when $x_f<0$, it is in the EMC region. Since
the parton distributions are evaluated at $m_\Upsilon^2$
and the gluon distribution evolves faster than the quarks,
we expect $S_A^\Upsilon > S_A^\psi$ for $x_f>0$.
In 1(c) and 1(d), the $x_f$ range corresponds to $-0.35<y<2.5$ and $|y|<1$,
the rapidity coverage of the PHENIX \cite{CDR} and ALICE \cite{ALICE}
detectors at RHIC
($\sqrt{s}=200A$ GeV) and the LHC ($\sqrt{s}=5.5$ TeV), respectively.
At RHIC central rapidities the \up\ may still be antishadowed, but
would be detected in the shadowing region with the muon spectrometer
\cite{CDR}.  At the LHC, both \jp\ and \up\ production
are shadowed.

The shadowing and antishadowing features of $S_A$ are shown explicitly in
Fig.\ 2 where we compare $S_A^\psi$ and $S_A^\Upsilon$ to the E772 data at
800 GeV on C, Ca, Fe, and W targets \cite{E772q}.  Recall that when $x_f$ is
increasing, $x_2$ is decreasing.  The solid curves show $S_A^\psi$, the
dashed, $S_A^{\psi^\prime}$, and the short dashed, $S_A^\Upsilon$.  The
difference between $J/\psi$ and $\psi^\prime$ comes from their
small mass difference.
Antishadowing is clearly seen for the $J/\psi$ as $x_f \rightarrow
0$.  Antishadowing at negative $x_f$ is in contradiction with the data:
clearly other effects are needed to account for this behavior.
The effect of $Q^2$ evolution on $S_A^\Upsilon$ is apparent--the antishadowing
peak is broader and closer to unity than the corresponding
region of $S_A^\psi$.  The \up\ enters the EMC region for $x_f < 0$, decreasing
$S_A^\Upsilon$.  While this trend agrees with the shape of the data,
the magnitude is too large.

Accounting for the primordial transverse momentum of the partons, $q_T$,
may reduce the magnitude
of $S_A$ since the transverse momentum spectrum may be
broadened in a nucleus because of Fermi motion.
If the cross sections are integrated over all $p_T$, the $q_T$ smearing
will have no effect (the $q_T$ spectrum in a moving nucleon
has the same normalization as for a stationary one).
However, the
data is taken over a finite range of $p_T$, suggesting
that the broader $q_T$ smearing in the nucleus manifests itself
in a reduced $p_T$-integrated cross section relative to the
nucleon, reducing $S_A$.
We calculated the $q_T$ spectra in nuclei by smearing the
primordial transverse
momentum distribution in a free nucleon with nucleon momentum
distributions from realistic many-body calculations \cite{momdist}.
The primordial $q_T$ spectrum for a free nucleon was
parametrized by a gaussian with
$\langle q_T^2 \rangle \approx 1 {\rm GeV^2}$, the same form
used to describe the low $p_T$ Drell-Yan data
in $pp$ scattering \cite{Antreasyan}. Similar values of
$\langle q_T^2 \rangle$ are also compatible with the low
$p_T$ data for \jp\ production \cite{Bad}.
The convolution formula also involves longitudinal degrees of freedom and
is therefore $x$ dependent, producing
an $x_f$-dependent effect on $S_A$.
A detailed description of our calculation will be
shown in a more extended paper \cite{RamSim}.
The nuclear $q_T$ smearing
shown in Fig.\ 2(b) and in Fig.\ 2(d) for the \jp\ (dotted curves)
and in Fig.\ 2(d) for \up\ (dot-dashed curve), reduces
the effect of antishadowing and affects the shadowing
part only marginally because of its increasing behavior
with $x$ (decreasing with $x_f$).
The agreement with data is substantially improved.

In Fig.\ 3(a) we compare the model with the E772 Drell-Yan production data
\cite{E772} in the W target as a check on the role of sea
quark shadowing
and the effect of $Q^2$ evolution.  The model, including
a small effect from $q_T$ smearing (dotted curve),
indeed gives a good description of the data.
We show a comparison with the NA3 \jp\ production data
from 200 GeV proton interactions
with $^2$H and Pt targets \cite{Bad} in 3(b).
The antishadowing region is
broad with shadowing only for $x_f > 0.4$.
The width of the antishadowing peak decreases with increasing energy since
$x_2$ passes from the shadowing region to the EMC regime over a smaller
range of $x_f$.  The antishadowing region is also shifted backward with
$x_f$ as the energy grows, indicated by a comparison of $S_A^\psi$ in Fig.\
2(d) and Fig.\ 3(b).
Clearly nuclear shadowing alone,
and even including the $p_T-$integrated spectra,
does not account for the observed suppression.
Other effects become important and must be taken into account
\cite{VBH1}.

Finally, we predict the shadowing of \jp, $\Upsilon$, and Drell-Yan
production in Au+Au collisions at RHIC energy
relative to $pp$ production at the same energy in Fig.\ 3(c).
Both the projectile and target
distributions are shadowed.  RHIC
data will not be strongly affected by antishadowing,
as expected from Fig.\ 1(c). Since only relatively
small $x_f$ values can be measured, features important
at large $x_f$ will not play a role.
Thus shadowing will dominate the nuclear effects.
We expect this behavior to be even more pronounced  for LHC,
where the range in $x$ extends below $x=10^{-3}$, Fig.\ 1(d).
Here recombination effects should play an important role and
could be tested accurately. Calculations including recombination effects
at LHC energies are in progress and will be shown elsewhere \cite{RamSim}.

We have evaluated the effect of shadowing on heavy quarkonium
production in current experiments and at RHIC.
Because of their difference in mass, $J/\psi$ and $\Upsilon$
production follow different patterns.
Antishadowing has been
shown to be substantial at present energies, in particular, \up\ production
is within the antishadowing region. However the
amount of shadowing and antishadowing
for $\Upsilon$ is smaller than for the $J/\psi$ because of
the perturbative $Q^2$ evolution of the shadowed distributions.
Nuclear modifications of the primordial transverse momentum spectrum are
an important addition in the low $p_T$ range.

Nuclear shadowing is an important feature of
quarkonium production, yet at lower energies it is not sufficient.
Shadowing will play a major role in
nucleus-nucleus collisions, suppressing both $J/\psi$
and $\Upsilon$ production significantly but according to different
patterns in $x_f$, over an
extrapolation from $pp$ interactions.

We thank K. J. Eskola, L. Frankfurt, S. Gavin, I. Hinchcliffe,
B. Kopeliovich, K.B. Luk, J. Peng, H. Schellman, M. Strikman and
X.-N. Wang for useful discussions.
S.L. would like to thank the Nuclear
Theory Group at the Lawrence Berkeley Laboratory for their warm
hospitality.

\newpage
\begin{center}
{\bf Figure Captions}
\end{center}
\vspace{0.5in}

\noindent Figure 1.  A comparison of $x_2$ as a function of $x_f$ for $J/\psi$
(solid) and $\Upsilon$ (dashed) production, at (a) 200 GeV and (b) 800 GeV
fixed target energies and (c) RHIC and (d) LHC ion collider energies.
\vspace{0.2in}

\noindent Figure 2.  Our model calculations for $S_A^\psi$ (solid),
$S_A^{\psi^\prime}$ (dashed), and $S_A^\Upsilon$ (dotted) compared to data
from (a) C, (b) Ca, (c) Fe, and (d) W targets at 800 GeV \cite{E772}.  The
$\Upsilon$ data is shown on (d) only.  The effect of
$q_T$ broadening in the nucleus is shown in (b) and (d) for
$S_A^\psi$ (dotted) and in (d) for $\Upsilon$ (dot-dashed).
\vspace{0.2in}

\noindent Figure 3.  (a) A calculation of $S_A^{DY}$ with the data from Ref.\
\cite{E772q} at 800 GeV on the W target (solid). The dotted curve
shows the effect of $q_T$ broadening. (b)  The $J/\psi$ production data
of Ref.\ \cite{Bad} on $^2$H and Pt targets at 200 GeV compared with
$S_A^\psi$.  (c) Predictions of $S_{AA}^\psi$ (solid), $S_{AA}^{\psi^\prime}$
(dashed),
$S_{AA}^\Upsilon$ (short-dashed),
and $S_{AA}^{DY}$ (dotted) for Au+Au collisions at RHIC.
\end{document}